\documentclass[12pt,epsfig,prd,showpacs]{revtex4}
\usepackage{graphicx}
\usepackage{epsfig}
\usepackage{dcolumn}
\usepackage{bm}

\begin{document}

\title{\boldmath Improved measurement of $\psi (2S)$ decays into $\tau
^{+}\tau ^{-}$} \author{M.~Ablikim$^{1}$, J.~Z.~Bai$^{1}$,
Y.~Ban$^{12}$, J.~G.~Bian$^{1}$, X.~Cai$^{1}$, H.~F.~Chen$^{17}$,
H.~S.~Chen$^{1}$, H.~X.~Chen$^{1}$, J.~C.~Chen$^{1}$, Jin~Chen$^{1}$
% ...
}

\affiliation{\vspace{0.2cm} (BES Collaboration)\\
\vspace{0.2cm} \textit{$^{1}$ Institute of High Energy Physics, Beijing
100049, People's Republic of China}\\
$^{2}$ China Center for Advanced Science and Technology (CCAST), Beijing
100080, People's Republic of China\\
$^{3}$ Guangxi Normal University, Guilin 541004, People's Republic of China\\
$^{4}$ Guangxi University, Nanning 530004, People's Republic of China\\
$^{5}$ Henan Normal University, Xinxiang 453002, People's Republic of China\\
$^{6}$ Huazhong Normal University, Wuhan 430079, People's Republic of China\\
$^{7}$ Hunan University, Changsha 410082, People's Republic of China\\
$^{8}$ Jinan University, Jinan 250022, People's Republic of China\\
$^{9}$ Liaoning University, Shenyang 110036, People's Republic of China\\
$^{10}$ Nanjing Normal University, Nanjing 210097, People's Republic of China\\
$^{11}$ Nankai University, Tianjin 300071, People's Republic of China\\
$^{12}$ Peking University, Beijing 100871, People's Republic of China\\
$^{13}$ Shandong University, Jinan 250100, People's Republic of China\\
$^{14}$ Sichuan University, Chengdu 610064, People's Republic of China\\
$^{15}$ Tsinghua University, Beijing 100084, People's Republic of China\\
$^{16}$ University of Hawaii, Honolulu, HI 96822, USA\\
$^{17}$ University of Science and Technology of China, Hefei 230026,
People's Republic of China\\
$^{18}$ Wuhan University, Wuhan 430072, People's Republic of China\\
$^{19}$ Zhejiang University, Hangzhou 310028, People's Republic of China\\
\vspace{0.2cm} $^{a}$ Current address: Purdue University, West Lafayette, IN
47907, USA\\
$^{b}$ Current address: DESY, D-22607, Hamburg, Germany\\
$^{c}$ Current address: Laboratoire de l'Acc{\'e}l{\'e}rateur Lin{\'e}aire,
Orsay, F-91898, France\\
$^{d}$ Current address: University of Michigan, Ann Arbor, MI 48109, USA}

\date{\today }

\begin{abstract}

Using $14$M $\psi (2S)$ events collected at BESII, the branching
fraction of $\psi (2S)\rightarrow \tau ^{+}\tau ^{-}$ is measured to
be $Br_{\tau \tau }=(3.10\pm 0.21\pm 0.38)\times 10^{-3}$, where the
first error is statistical and the second is systematic.

\end{abstract}

 \maketitle

\section{\boldmath Introduction}

The decays $\psi (2S)\rightarrow l^{+}l^{-}$ ($l=e,$ $\mu$, or $\tau$ )
have been measured by E760~\cite{fermilab1}, E835~\cite{fermilab2},
BES~\cite{bes}, and BaBar~\cite{babar}, and branching fractions are
listed in Table~\ref{measured}. According to the sequential lepton
hypothesis, the branching fractions of these decays satisfy \cite{pdg}

$$
Br_{ee}\simeq Br_{\mu \mu }\simeq \frac{Br_{\tau \tau }}{0.3885}\equiv
Br_{ll},
$$
and the measurements agree with this relation within the large uncertainties.

The decay $\psi (2S)\rightarrow \tau ^{+}\tau ^{-}$ was first observed by
DASP~\cite{dasp}, and the branching fraction was first measured by
BESI using $ 4\times 10^{6}$ $\psi(2S)$ events. In
this paper, we report an improved measurement of the branching
fraction for $\psi (2S)\rightarrow \tau ^{+}\tau ^{-}$. The
measurement is based on a data sample of $14(1\pm 4\%)\times 10^{6}$
$\psi (2S)$ events collected by the BESII detector at the BEPC.

\begin{table}[h]
\caption{Experimental results for $Br(\psi(2S)\rightarrow l^{+}l^{-})$ $(\times 10^{-3})$. The  CLEO-c result
is calculated from $\Gamma _{ee}$\cite{cleo}$/\Gamma_{\psi(2S)}$\cite{pdg}.}
\label{measured}
\begin{tabular}{lcccc}
\hline\hline
Experiment & Year & $Br_{ee}$ & $Br_{\mu\mu}$ & $Br_{\tau\tau}$ \\ \hline
E760~\cite{fermilab1} & 1996 & $8.3\pm 0.86$ & - & - \\
E835~\cite{fermilab2} & 2000 & $7.4\pm 0.73$ & - & - \\
BES~\cite{bes} & 2000 & - & - & $2.71\pm 0.7$ \\
BaBar~\cite{babar} & 2003 & $7.8\pm 1.2$ & $6.7\pm 1.1$ & - \\
CLEO-c & 2006 & $7.5\pm 0.3$ & - & - \\ \hline
PDG~\cite{pdg} & 2006 & $7.35\pm 0.18$ & $7.3\pm 0.8$ & $2.8\pm 0.7$ \\ \hline\hline
\end{tabular}
\end{table}

\section{\boldmath BES detector}

The upgraded Beijing Spectrometer (BESII) is located at the Beijing
Electron-Positron Collider (BEPC). BESII is a large solid-angle magnetic
spectrometer which is described in detail in Refs. \cite{bes2,bes2'}. The
momentum of the charged particle is determined by a 40-layer cylindrical
main drift chamber (MDC) which has a momentum resolution of $\sigma
_{p}/p=1.78\%\sqrt{1+p^{2}}$ ($p$ in GeV/$c$). Particle identification is
accomplished by specific ionization ($d$E$/dx$) measurements in the drift
chamber and time-of-flight (TOF) information in a barrel-like array of $48$
scintillation counters. The $d$E$/dx$ resolution for hadron tracks is about
8.0\%; the TOF resolution\ of charged hadrons is about $200$ ps. Radially
outside of the TOF counters is a $12$-radiation-length barrel shower counter
(BSC) comprised of gas tubes interleaved with lead sheets. The BSC measures
the direction and energy of photons with an energy resolution of $\sigma _{E}/
E \approx 21\%/\sqrt{E}$ ($E$ in GeV). Outside of the solenoidal coil,
which provides a $0.4$\ Tesla magnetic field over the tracking volume, is an
iron flux return that is instrumented with three double layers of counters
that identify muons of momentum greater than $0.5$ GeV/$c$. It provides
coordinate measurements along the muon trajectories with resolutions in the
outermost layer of $10$ cm in $r\phi$ and $12$ cm in $z$. The solid
angle coverage of the layers is $67\%$, $67\%$, and $63\%$ of $4\pi $,
respectively.

A GEANT3 based Monte Carlo (MC) program with detailed consideration of the
detector performance is used \cite{simbes}. Reasonable agreement between
data and Monte Carlo simulation has been observed in various channels
tested, including $e^{+}e^{-}\rightarrow (\gamma )e^{+}e^{-}$, $
e^{+}e^{-}\rightarrow (\gamma )\mu ^{+}\mu ^{-}$, $J/\psi \rightarrow p
\overline{p}$, and $\psi (2S)\rightarrow \pi ^{+}\pi ^{-}J/\psi $, $J/\psi
\rightarrow l^{+}l^{-}(l=e,\mu )$.

\section{\boldmath Event Selection}

The $\tau $ pair events are identified by the topology $\psi
(2S)\rightarrow \tau ^{+}\tau ^{-}\rightarrow e\nu \overline{\nu }\mu
\nu \overline{\nu }$ (denoted as $e\mu $). These events are
characterized  by two charged tracks, missing energy and momentum, and
no real hits in the BSC.  An event is required to
have two oppositely charged tracks, each of which is well fitted to a
helix within the polar angle region $\left\vert \cos \theta
\right\vert <0.8$ and with the point of closest approach of the track
to the beam line within the interaction region of
$\sqrt{x_{0}^{2}+y_{0}^{2}}<2$ cm, $\left\vert z_{0}\right\vert <20 $
cm. The transverse momentum of each track is required to be greater
than $70$ MeV/$c$, which is the minimum needed to reach the outer
radius of the BSC in the $0.4$ Tesla magnetic field, and the momentum
is required to be less than $1.2 $ GeV/$c$ to reject background from
radiative Bhabha ($e^+ e^- \to (\gamma) e^+ e^-$) and dimuon ($e^+ e^-
\to (\gamma) \mu^+ \mu^-$) events.

The electron is identified using the following selection criteria. The
ratio of the energy deposited by the track in the BSC to its momentum
($E/p$) should be greater than $0.65$. To further distinguish the
electron from hadrons, we determine
$$X_{se}=\frac{(\frac{dE}{dx})_{meas}-(\frac{dE}{dx})_{exp}}{\sigma_{\frac{dE}{dx}}}$$
and
$$T_{se}=\frac{(TOF)_{meas}-(TOF)_{exp}}{\sigma_{TOF}},$$ where
$(\frac{dE}{dx})_{meas}$ and $(TOF)_{meas}$ are the measured $dE/dx$
and TOF values, $(\frac{dE}{dx})_{exp}$ and $(TOF)_{exp}$ are the
expected values for the electron hypothesis , and
$\sigma_{\frac{dE}{dx}}$ and $\sigma_{TOF}$ are the resolutions.  A
radiative Bhabha sample is used to determine corrections so that
$X_{se}$ and $T_{se}$ are standard normal distributions. We require
$\left\vert X_{se}\right\vert <2.5$ and $\left\vert T_{se}\right\vert
<2.5$. The combined $X_{se}$ and $T_{se}$ information is used to
calculate the confidence level for the electron hypothesis, and it is
required to be greater than $0.01$.

To identify the muon, the difference between the closest muon hit
position in the muon counter and the projected MDC track in the $i$-th
layer ($i=1,$ $2,$ $3$), $\delta _{i}$, is used. A hit is considered
as a good $\mu$ hit if $\delta _{i}$ is within three standard
deviations, where the $\mu^{+}$ and $\mu^{-}$ $\delta _{i}$ distributions as a function of momentum and  $\cos \theta$
 have been corrected to standard normal distributions using radiative dimuon events.  The total number of the good $\mu $ hits, $\mu
_{hit}^{good}$, should be greater than one.

The total energy of the neutral clusters in the BSC, which are not
associated with the charged tracks, $E_{neu}^{tot}$, is required to be
less than $0.2$ GeV to remove the backgrounds which contain pions or gammas.

\section{\boldmath Efficiencies}

The $e$ and $\mu $ identification efficiencies, $\varepsilon _{eID}$
and $\varepsilon _{\mu ID}$, are determined using radiative Bhabha and dimuon events, and the remaining
efficiency $\varepsilon _{MC}$, including geometric acceptance, is
determined from Monte Carlo simulation. Radiative Bhabha and dimuon events require that the higher momentum track in each event be a well identified electron or muon. The overall
efficiency is $\varepsilon _{e\mu }=\varepsilon _{eID}\times
\varepsilon _{\mu ID}\times \varepsilon _{MC}$.

Radiative Bhabha events are used to measure $\varepsilon _{eID}$.  It
is $ (82.8\pm 0.08)\%$, while from MC simulation, it is $(81.5\pm
0.11)\%$; the difference $0.2\%$ is take as the systematic error for
$e$ identification.

Radiative dimuon events are used to measure $\varepsilon _{\mu
ID}$. The $\mu$ identification efficiencies vary with the $\mu $
transverse momentum region, so $\varepsilon _{\mu ID} =
\sum_{i}\omega _{i}\varepsilon _{i}^{\mu }$, where
$\varepsilon _{i}^{\mu }$ is the $\mu $ identification efficiency in
the $i$-th transverse momentum bin determined from data, and $\omega
_{i}$ is the fraction of MC simulated $\psi (2S)\rightarrow \tau
^{+}\tau ^{-}\rightarrow e\nu \overline{\nu }\mu \nu \overline{\nu }$
events in the same bin.  Table \ref{usys2} shows $\omega _{i}$ and
$\varepsilon _{i}^{\mu }$; $\varepsilon _{\mu ID}$ is determined to be
$33.9(1\pm 0.022)\%$.  The final efficiency for selecting $e\mu $
events is $\varepsilon _{e\mu }=17.8\%.$

\begin{table}[tbp]
\caption{The $\omega _{i}$ and $\varepsilon _{i}^{\mu }$ values in different $P_{xy}$ regions.}
\label{usys2}$
\begin{array}{ccc}
\hline\hline
P_{xy}^{\mu }\text{ (GeV/}c\text{)} & \omega _{i}(\%) & \varepsilon
_{i}^{\mu }(\%) \\ \hline
0.5<P_{xy}<0.6 & 15.45 & 55.0\pm 5.3 \\
0.6<P_{xy}<0.7 & 13.69 & 66.7\pm 2.4 \\
0.7<P_{xy}<0.8 & 10.02 & 77.2\pm 1.2 \\
0.8<P_{xy}<0.9 & 6.08 & 81.7\pm 0.89 \\
0.9<P_{xy}<1.0 & 3.11 & 83.3\pm 0.66 \\
1.0<P_{xy}<1.1 & 1.07 & 85.7\pm 0.45 \\
1.1<P_{xy}<1.2 & 0.11 & 88.5\pm 0.31 \\ \hline\hline
\end{array}
$
\end{table}

\section{\boldmath Background Estimation}

In this experiment, $\tau \tau $ pairs are produced by: (1) $\psi
(2S)$ decays, (2) direct continuum production, and (3) the
interference between them. To measure the branching fraction of
$\psi(2S)\rightarrow \tau ^{+}\tau ^{-}$, the continuum contribution
including interference must be subtracted. Other $\psi
(2S)$ decay backgrounds must also be subtracted.

The backgrounds from $\psi (2S)$ decays are studied using the 14 M MC
inclusive $\psi (2S)$ decays generated with Lundcharm \cite{lund}. The
dominant background is found to be from $\psi (2S)\rightarrow \tau
^{+}\tau ^{-}\rightarrow e\nu \overline{\nu }\pi \nu $. All possible
two-body background channels are generated according to the branching
fractions given in PDG(2006), including $\psi (2S)\rightarrow \tau
^{+}\tau ^{-}$ with $\tau \to \pi \nu $ or $ \to \pi \pi ^{0}\nu$.
Using $Br_{\tau \tau }=3.10\times 10^{-3}$, which is the value
determined in this experiment, the number of the background events
from $\psi(2S)$ decays is $N_{bg}^{norm}=68.5\pm 3.2$.

The continuum background is estimated by applying the same selection
criteria to the data sample taken at $\sqrt{s}=3.650$ GeV and
normalizing the result to
$3.686$ GeV:
$$
N_{cont}^{obs}=N_{3.650}^{obs}\cdot \frac{\sigma _{3.686}^{cont}}{\sigma
_{3.650}^{cont}}\cdot \frac{L_{3.686}}{L_{3.650}}.
$$
where $N$ is the number of continuum background events, $\sigma ^{cont}$ is
the cross section for the continuum process, and $L$ is the integrated
luminosity.  The continuum background at $\sqrt{s}=3.686$ GeV is
estimated to be $N_{cont}^{obs}=516.4\pm 45.3$.

\section{\boldmath Branching fraction of $\protect\psi (2S)\rightarrow
\protect\tau ^{+}\protect\tau ^{-}$}

The branching ratio $Br_{\tau \tau }$ is determined from:

\begin{equation}
Br_{\tau \tau }=\frac{\frac{N^{obs}-N_{cont}^{obs}-N_{bg}^{norm}(Br_{\tau
\tau })}{\varepsilon _{e\mu }\cdot Br(e\mu )}-\sigma _{Int}^{\tau \tau
}(Br_{\tau \tau })\cdot L_{3.686}}{N_{\psi (2S)}},  \label{Br4}
\end{equation}

\noindent where $N^{obs}$ is the number of observed events, $Br(e\mu)$ is the
fraction of $\tau ^{+}\tau ^{-}$ events producing the $e\mu$ topology,
which is $2Br(\tau \rightarrow e\nu \widetilde{\nu } )Br(\tau
\rightarrow \mu \nu \widetilde{\nu })=0.06194$ \cite{pdg}, $L_{3.686}$
is the integrated luminosity of the $\psi(2S)$ data, $N_{\psi (2S)}$
is the total number of $\psi(2S)$ events, and $\sigma _{Int}^{\tau
\tau }(Br_{\tau \tau })$ is the interference cross section between
$\psi(2S)$ decay and continuum amplitudes.  A summary of the numbers
used to  find $Br_{\tau \tau}$  is given in Table~\ref{resultsummary}.

Since $N_{bg}^{norm}(Br_{\tau \tau })$ and
$\sigma _{Int}^{\tau \tau }(Br_{\tau \tau })$ depend on $Br_{\tau \tau
}$, we can solve Eq. \ref{Br4} to obtain $Br_{\tau \tau }$. The value of
$N_{bg}^{norm}(Br_{\tau \tau })$ is $22086.8\cdot Br_{\tau \tau
}$. Substituting the total width of $\psi(2S)$, $\Gamma =281$ keV
\cite{pdg}, and the energy spread $\Delta _{E}=1.3$ MeV~\cite{beam} into the
cross section function given in Ref.~\cite{equation} , $\sigma _{Int}^{\tau
\tau }(Br_{\tau \tau })=-66.587\cdot Br_{\tau \tau }$ is obtained at
$\sqrt{s}=3.686$ GeV. Solving, $Br_{\tau \tau }$ is $3.10\times 10^{-3}$.

\begin{table}[h]
\caption{Summary of numbers used to determine $Br_{\tau \tau }$.}
\label{resultsummary}
\begin{center}
\begin{tabular}{lc}
\hline\hline
$N^{obs}$              & 1015   \\
$N_{cont}^{obs}$       &  $516.4\pm45.3$ \\
$\varepsilon _{e\mu }$ & 17.8\% \\
$Br(e\mu)$             & 0.06194 \\
$L_{3.686}$           & $19.72\pm 0.86$ pb$^{-1}$\\
$N_{\psi (2S)}$        & $14(1\pm 0.04)\times 10^{6}$ \\
$N_{bg}^{norm}(Br_{\tau\tau })$ & $68.5\pm 3.2$ \\
\hline
\end{tabular}
\end{center}
\end{table}

\begin{figure}[tbp]
\includegraphics[width=\columnwidth]{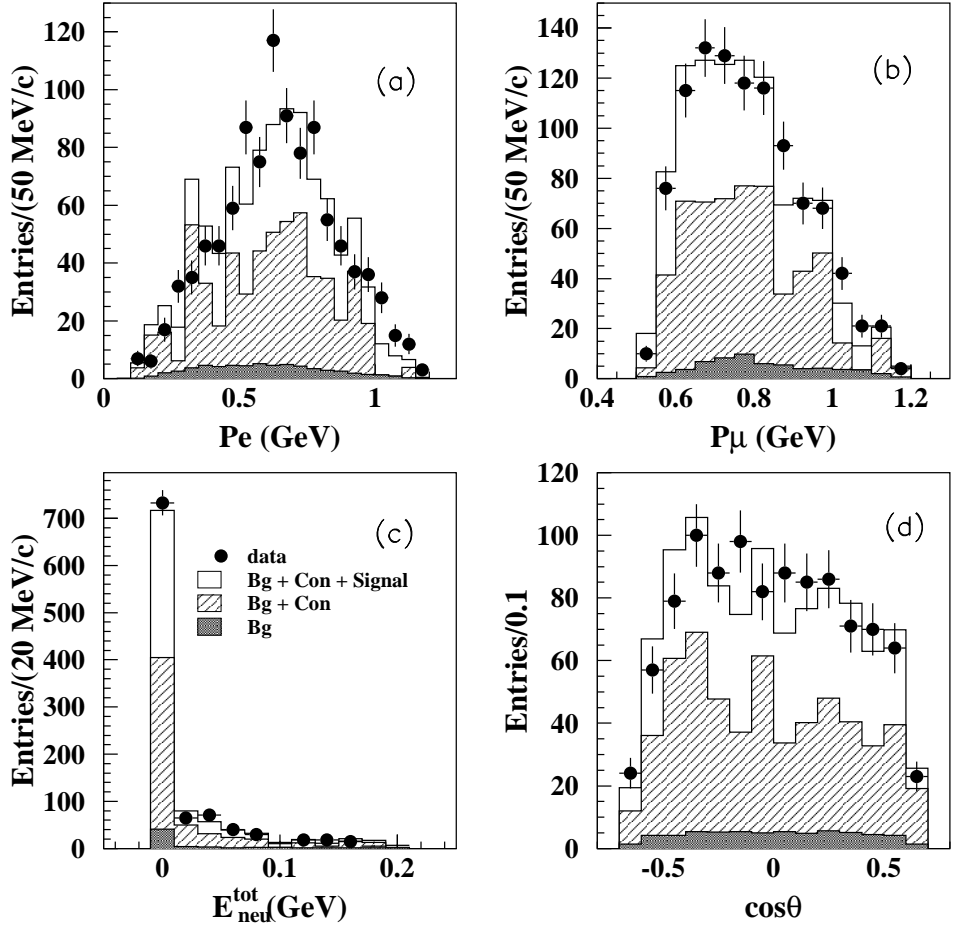}
\caption{Distributions for (a) the electron momentum, (b) muon momentum,
(c) $E_{neu}^{tot}$, and (d) cosine of the acollinearity angle. Dots with
error bars are data, the shaded histograms are the
background from $\psi (2S)$ decays, the histograms filled
with diagonal lines are the background from the continuum and from
$\psi(2S)$ decays, and the blank histograms are the signal plus the backgrounds.
}
\label{p_emu}
\end{figure}

Figure \ref{p_emu} shows distributions of (a) the electron momentum,
(b) the muon momentum, (c) $E_{neu}^{tot}$, and (d) the cosine of the
acollinearity angle.  The data (dots with error bars) are consistent
with Monte Carlo simulation (blank histogram), which includes
signal and backgrounds from $\psi(2S)$ decays and the continuum.

\section{\boldmath Systematic Errors}

\subsection{\boldmath Systematic error on $\varepsilon _{e\mu }$}

The systematic error on $\varepsilon _{e\mu }$ comes mainly from the
particle identification efficiency uncertainties, the MDC tracking
uncertainty, and the uncertainty of the $E_{neu}^{tot}$ requirement. As
shown in Sect. IV, the systematic errors for $e$ and $\mu $
identification are $0.2\%$ and $2.2\%$, respectively. The simulation
of the MDC tracking efficiency agrees with data within 1 to 2 \% for
each charged track as measured using channels like $J/\psi \rightarrow
\Lambda \overline{\Lambda }$ and $\psi (2S)\rightarrow \pi ^{+}\pi
^{-}J/\psi $, $J/\psi \rightarrow \mu ^{+}\mu ^{-}$. The systematic
error for the channel of interest is taken conservatively as $4\%$.

The effect of the requirement on the total energy of the neutral
clusters is estimated using the following method.  A neutral cluster
in the BSC is considered to be a photon candidate when the angle
between the nearest charged track and the cluster in the $xy$ plane is
greater than $15^{\circ }$, the first hit layer is in the beginning 6
radiation lengths, and the angle between the cluster development
direction in the BSC and the photon emission direction in the $xy$
plane is less than $37^{\circ }$. Instead of the requirement on the
energy in the BSC, we require that no photon is reconstructed. The
efficiency difference between data and MC for the no photon
requirement is measured to be $0.9\%$,
which we take as the systematic error for the total energy of neutral
clusters requirement.  Based on the above, $\varepsilon _{e\mu }$ is
$17.8(1\pm 4.7\%)\%$.

\subsection{\boldmath Systematic error of $\sigma _{inf}^{\tau \tau}$}

The systematic error on $\sigma _{inf}^{\tau \tau }$ comes from a
shift of the beam energy during data taking. About one third of the
$\psi (2S)$ data was acquired with the beam energy dropping slowly
from 1.8430 GeV to 1.8425 GeV. Defining $\sigma _{inf}^{\prime
}=\left\vert (\sigma _{inf}^{3.6855}/3+2\sigma
_{inf}^{3.686}/3)\right\vert$, the difference between $\sigma
_{inf}^{\prime }$ and $\sigma _{inf}^{\tau \tau }$ is $5.1\%$ and is
considered as the systematic error.

\subsection{\boldmath Systematic error of the branching fraction}

Table \ref{systemerror} shows the summary of the systematic errors,
where $\sigma _{abs}$ is the absolute uncertainty, and $\sigma _{rel}$
is the relative error. The total systematic error on the branching
fraction is $12.3\%$.
\begin{table}[h]
\caption{Summary of systematic errors.}
\label{systemerror}
\begin{center}
\begin{tabular}{lcc}
\hline\hline
source & $\sigma _{abs}$ & $\sigma _{rel}$ (\%)\\ \hline
$\varepsilon _{e\mu }$ & $0.12$ & $4.0$ \\
$\sigma _{Int}$ & $0.15$ & $5.0$ \\
$Br_{e\mu }$ & $0.013$ & $0.42$ \\
$L_{3.686}$ & $0.013$ & $0.42$ \\
$N_{\psi (2S)}$ & $0.13$ & $4.2$ \\
$N_{cont}^{obs}$ & $0.30$ & $10.0$ \\ \hline
Total & $0.38$ & $12.3$ \\ \hline\hline
\end{tabular}
\end{center}
\end{table}

\section{\boldmath Summary}

The process $\psi (2S)\rightarrow \tau ^{+}\tau ^{-}$ is studied with
$14\times 10^{6}$ $\psi (2S)$ events. The branching fraction is
measured to be $(3.10\pm 0.21_{stat.}\pm 0.38_{sys.})\times
10^{-3}$. The error is more precise than the previous measurement by BESI
\cite{bes}. In the present measurement, the $e$ and $\mu $ particle
identification efficiencies and the continuum background are obtained
from data.  The biggest systematic error comes from the
the continuum background. To obtain more precision, a
larger  continuum sample is required.

{\normalsize \acknowledgments}

The BES collaboration thanks the staff of BEPC for their hard efforts. This
work is supported in part by the National Natural Science Foundation of
China under contracts Nos. 10491300, 10225524, 10225525, the Chinese Academy
of Sciences under contract No. KJ 95T-03, the 100 Talents Program of CAS
under Contract Nos. U-11, U-24, U-25, and the Knowledge Innovation Project
of CAS under Contract Nos. U-602, U-34 (IHEP); by the National Natural
Science Foundation of China under Contract No. 10175060 (USTC), and No.
10225522 (Tsinghua University); and by the Department of Energy under
Contract No. DE-FG02-04ER41291 (University of Hawaii).

\end{document}